\documentclass[prb,preprint,groupedaddress, showkeys,showpacs]{revtex4}
\usepackage{graphicx}
\begin{document}

%\preprint{}

\title{A channel Brownian pump powered by an unbiased external force}


\author{Bao-quan  Ai$^{1}$}\email[Email: ]{aibq@hotmail.com}
\author{Liang-gang Liu$^{2}$}
%\homepage[]{}

%\thanks{}
\affiliation{$^{1}$ Institute for Condensed Matter Physics, School
of Physics and Telecommunication Engineering and Laboratory of
Photonic Information Technology, South China
Normal University, 510006 Guangzhou, China\\
$^{2}$ Faculty of Information Technology , Macau University of
Science and Technology, Macao}


\date{\today}
\begin{abstract}
\indent A Brownian pump of particles in an asymmetric finite tube is
investigated  in the  presence of an unbiased external force. The
pumping system is bounded by two particle reservoirs. It is found
that the particles can be pumped through the tube from a reservoir
at low concentration to one at the same or higher concentration.
There exists an optimized value of temperature (or the amplitude of
the external force) at which the pumping capacity takes its maximum
value. The pumping capacity decreases with increasing the radius at
the bottleneck of the tube.

\end{abstract}

\pacs{05. 06. Cd, 02. 50. Ey, 05. 40. Jc, 66. 10. Cb }



\maketitle


\section {Introduction}
\indent Noisy transport far from equilibrium play a crucial role
in many processes from physical and biological to social systems.
Molecular motors are the paradigm of how to extract useful
mechanical energy in the random environment of thermal
fluctuations \cite{1,2,3}. Inspired on Feynman's ratchet and pawl
machine, the appearance of directed motion enhanced by thermal
fluctuations has been named after the ratchet effect \cite{4}.

\indent The idea of applying the ratchet mechanism to model pumps
has already appeared in the literature \cite{5,6,7,8,9,10,11}. Prost
and co-workers \cite{5} studied the transport of an asymmetric pump
with a simple two-level model and quantified how vectorial symmetry
plus dissipation creates a macroscopic motion, even in the absence
of any externally applied gradient.  Astumian and Derenyi \cite{6}
investigated a chemically driven molecular electron pump in which
charge can be pumped through a tiny gated portal from a reservoir at
low electrochemical potential to on at the same or higher
electrochemical potential by cyclically modulating the portal and
gate energies. Kosztin and Schulten \cite{7} studied the
fluctuation-driven molecular transport through an asymmetric
potential pump and three transport mechanisms: driven by potential
gradient, by an external periodic force and by nonequilibrium
fluctuations.  Nonadiabatic electron heat pump was investigated by
Rey and coworkers \cite{8}. They presented a mechanism for
extracting heat metallic conductors based on the energy-selective
transmission of electrons through a spatially asymmetric resonant
structure subject to ac driving. Wambaugh and co-workers \cite{9}
studied the transport of fluxons in superconductors by alternating
current rectification. They found that at nonzero temperatures, the
asymmetric geometry of the ratchet sawteeth automatically converts
applied ac inputs into a net dc motion of fluxons. In their another
work \cite{10}, they studied how inertia and particle pair
interactions can induce transverse rectification of particle flows
driven through square arrays or chains of triangular defects
perpendicularly to their symmetry axis.  Recently, Sancho and
Gomez-marin \cite{11} presented a model for a Brownian pump powered
by a flashing ratchet mechanism. The pumping device was embedded in
a finite region and bounded by particle reservoirs. Their emphasis
is on finding what concentration gradient the pump can maintain.

\indent The previous works on pump were on the consideration of the
energy barriers. The present work is extend to the study of the
molecular pump to the case of the entropic barriers. The entropic
barriers may appear when coarsening the description of a complex
system for simplifying its dynamics \cite{12,13,14}. We emphasize on
finding how the Brownian particles can be pumped through an
asymmetric tube from a particle reservoir at low concentration to
one at the same or higher concentration in the presence of an
unbiased external force. Our other task is on investigation what
factors can affect the pumping capacity.

\section {General analysis}

%\begin{figure}[htbp]
 % \begin{center}\includegraphics[width=8cm,height=4cm]{fig1.eps}
 %\caption{\baselineskip 0.4in Scheme of the pumping device: a spatially asymmetric tube is embedded in a finite region of length $L$ and bounded by two particle reservoirs of concentrations $\rho_{0}$ and $\rho_{1}$.
 %The shape of the tube is determined by its radius $\omega(x)$ and the left end coordinate $x_{0}$ of the tube. The particles in the tube are powered by an unbiased external force $F(t)$.}\label{1}
%\end{center}
%\end{figure}

\indent We consider overdamped Brownian particles moving in an
asymmetric finite tube [Fig. 1] in the presence of an unbiased
external force. The tube is embedded in a finite region and bounded
by two particle reservoirs. Its overdamped dynamics is described by
the following Langevin equations\cite{12,15}

\begin{equation}\label{}
    \gamma\frac{dx}{dt}=F(t)+\xi_{x}(t),
\end{equation}
\begin{equation}\label{}
    \gamma\frac{dy}{dt}=\xi_{y}(t),
\end{equation}
\begin{equation}\label{}
    \gamma\frac{dz}{dt}=\xi_{z}(t),
\end{equation}
where $x$, $y$, $z$  are the three-dimensional (3D) coordinates,
$\gamma$ is the friction coefficient of the particle.
$\xi_{x,y,z}(t)$ are the uncorrelated Gaussian white noises with
zero mean and correlation function:
$<\xi_{i}(t)\xi_{j}(t^{'})>=2\gamma
k_{B}T\delta_{i,j}\delta(t-t^{'})$ for $i,j=x, y, z$. $T$ is the
absolute temperature. $k_{B}$ is the Boltzmann constant.  $<...>$
denotes an ensemble average over the distribution of noise. Imposing
reflecting boundary conditions in the transverse direction ensures
the confinement of the dynamics within the tube. The shape of the
tube is described by its radius,
\begin{equation}\label{}
    \omega(x)=a\sin(\frac{2\pi x}{L})+b, x_{0}\leq x \leq
    x_{0}+L,
\end{equation}
where $a$ is the parameter that controls the flat degree of the tube
of the tube, $L$ is the length of the tube. The radius at the
bottleneck is $r_{b}=b-a$. $x_{0}$ is the coordinate of the left
end.  $F(t)$ is an unbiased external force and satisfies
\begin{equation}\label{}
 F(t)=\left\{
\begin{array}{ll}
   F_{0},& \hbox{$n\tau\leq
t<n\tau+\frac{1}{2}\tau$};\\
   -F_{0} ,&\hbox{$n\tau+\frac{1}{2}\tau<t\leq
      (n+1)\tau$},\\
\end{array}
\right.
\end{equation}
where $\tau$ is the period of the unbiased force and $F_{0}$ is its
magnitude.

\indent When the particles move in a confined media, its movement
equation can be described by the Fick-Jacobs equation
\cite{12,13,14,15} which is  derived from the 3D (or 2D)
Smoluchowski equation after elimination of $y$ and $z$ coordinates
by assuming equilibrium in the orthogonal directions. The
complicated boundary conditions of the diffusion equation in
irregular channels can be greatly simplified by introducing an
entropic potential that accounts for the reduced space accessible
for diffusion of the Brownian particle. Reduction of the coordinates
may involve the appearance of the entropic barriers and an effective
diffusion coefficient. When $|\omega^{'}(x)|<1$, the effective
diffusion coefficient reads \cite{12,13,14,15}
\begin{equation}\label{}
    D(x)=\frac{D_{0}}{[1+\omega^{'}(x)^{2}]^{\alpha}},
\end{equation}
where $D_{0}=k_{B}T/\gamma$ and $\alpha=1/2$ for three dimensions.
The prime stands for the derivative with respect to the space
variable $x$.

\indent Consider the effective diffusion coefficient and the
entropic barriers, the dynamics of a Brownian particle moving along
the axis of the 3D tube can be described by the equation
\cite{12,15}
\begin{equation}\label{}
    \frac{\partial \rho(x,t)}{\partial t}=\frac{\partial}{\partial x}[D(x)\frac{\partial \rho(x,t)}{\partial
    x}+\frac{D(x)}{k_{B}T}A^{'}(x,t)\rho(x,t)]=-\frac{\partial j(x,t)}{\partial
    x},
\end{equation}
where a free energy $A(x,t)=E-TS=-F(t)x-Tk_{B}\ln h(x)$ is defined
\cite{12,13,14}: here $E=-F(t)x$ is the energy, $S=k_{B}\ln h(x)$ is
the entropy, $h(x)$ is the dimensionless transverse cross section
$\pi[\omega(x)/L]^{2}$ of the tube in three dimensions. $j(x,t)$ is
the probability current and $\rho(x,t)$ is the particle
concentration.

\indent If $F(t)$ changes very slowly with respect to $t$, namely,
its period is longer than any other time scale of the system,
 there exists a quasistatic state. In the steady state, the
 concentration is just a function of space thus the flux becomes a constant $j$.
 The concentration $\rho(x)$ follows a first order non homogeneous linear
 differential equation, whose formal solution is
\begin{equation}\label{}
    \rho(x)=\exp[-\int_{x_{0}}^{x}\frac{A^{'}(z)}{k_{B}T}dz]\{c_{0}-j\int_{x_{0}}^{x}\frac{dz}{D(z)}\exp[\int_{x_{0}}^{z}\frac{A^{'}(y)}{k_{B}T}dy]\}.
\end{equation}
\indent Though, unlike typical Brownian motors, the boundary
conditions are not periodic nor the normalized condition is
imposed, the unknown constant $c_{0}$ can be found by imposing the
left reservoir concentration $\rho_{0}\equiv \rho(x_{0})$ and the
right concentration $\rho_{1}\equiv \rho(x_{0}+L)$ as fixed
boundary conditions \cite{11}. Then $c_{0}=\rho(x_{0})$ and
\begin{equation}\label{}
    j(F_{0})=\frac{k_{B}T[\rho_{0}-\rho_{1}e^{-\frac{F_{0}L}{k_{B}T}}]}{\int_{x_{0}}^{x_{0}+L}[1+\omega^{'}(x)^{2}]^{\alpha}e^{-\frac{F_{0}(x-x_{0})}{k_{B}T}}[\frac{\omega(x_{0})}{\omega(x)}]^{2}dx}.
\end{equation}
\indent The average current is
\begin{equation}\label{}
    J=\frac{1}{\tau}\int^{\tau}_{0}j(F(t))dt=\frac{1}{2}[j(F_{0})+j(-F_{0})].
\end{equation}

\indent For studying the pumping capacity, we consider the
situation in which $J$ tends to zero which corresponds the case in
which the pump in maintaining the maximum concentration difference
between the two reservoirs across the membrane with no net leaking
of particle. The method is the same that in Ref. 11. This
situation is analogous the stalling force in Brownian motors.

From Eqs. (8-10), we can obtain

\begin{equation}\label{}
    \frac{\rho_{1}}{\rho_{0}}=\frac{\int_{x_{0}}^{x_{0}+L}[1+\omega^{'}(x)^{2}]^{\alpha}[e^{\frac{F_{0}(x-x_{0})}{k_{B}T}}+e^{-\frac{F_{0}(x-x_{0})}{k_{B}T}}][\frac{\omega(x_{0})}{\omega(x)}]^{2}dx}{e^{-\frac{F_{0}L}{k_{B}T}}\int_{x_{0}}^{x_{0}+L}[1+\omega^{'}(x)^{2}]^{\alpha}e^{\frac{F_{0}(x-x_{0})}{k_{B}T}}[\frac{\omega(x_{0})}{\omega(x)}]^{2}dx+e^{\frac{F_{0}L}{k_{B}T}}\int_{x_{0}}^{x_{0}+L}[1+\omega^{'}(x)^{2}]^{\alpha}e^{-\frac{F_{0}(x-x_{0})}{k_{B}T}}[\frac{\omega(x_{0})}{\omega(x)}]^{2}dx}.
\end{equation}

\section {Results and discussions}

\indent For simplicity, we take $k_{B}=1$, $\gamma=1$ and $L=2\pi$
throughout the study. Our method is valid only when
$|\omega^{'}(x)|<1$ and its validity is testified by using numerical
simulations \cite{16}. Thus, we take $a=\frac{1}{2\pi}$ and
$|\omega^{'}(x)|<<1$ which ensure the validity of our method.

\subsection{Current $J$ at $\rho_{1}=\rho_{0}$}
%\begin{figure}[htbp]
 % \begin{center}\includegraphics[width=10cm,height=8cm]{fig2.eps}
%\caption{ Current $J$ as a function of period numbers $N$ at
%$x_{0}=0$, $a=\frac{1}{2\pi}$, $b=\frac{1.5}{2\pi}$, $\alpha=1/2$,
%$F_{0}=0.5$, $T=0.5$, $\rho_{0}=1$, and $\rho_{1}=1$. }\label{1}
%\end{center}
%\end{figure}

\indent In order to check the effect of the magnitude for the tube,
we plot the current $J$ as a function of period numbers $N$( we have
$N$ cells in the channel) in Fig. 2. It is found that the current
$J$ is independent of the period numbers $N$. Each cell has the same
structure and pumping capacity. Thus, our results are not finite
size effects.

%\begin{figure}[htbp]
 % \begin{center}\includegraphics[width=10cm,height=8cm]{fig3a.eps}
 % \includegraphics[width=10cm,height=8cm]{fig3b.eps}
%\caption{(a) Current $J$ as a function of $x_{0}$ for different
%values of $F_{0}$ at $a=\frac{1}{2\pi}$, $b=\frac{1.5}{2\pi}$,
%$\alpha=1/2$, $T=0.5$, $\rho_{0}=1$, and $\rho_{1}=1$. (b)Currents
%$j(F_{0})$, $-j(-F_{0})$ and $J$ versus $x_{0}$ for $F_{0}=0.5$. The
%other parameters are the same as that in (a) }\label{1}
%\end{center}
%\end{figure}

\indent Figure 3(a) shows the current as a function of the initial
coordinate $x_{0}$ for different values of $F_{0}$ at
$\rho_{0}=\rho_{1}$. In this case, the transport behavior is similar
to that in a periodic tube. The current is positive for $0\leq
x_{0}<\frac{\pi}{2}$ and $\frac{3}{2}\pi<x_{0}\leq 2\pi$, zero at
$x_{0}=\frac{\pi}{2}$ and $\frac{3}{2}\pi$, and negative for
$\frac{\pi}{2}<x_{0}<\frac{3}{2}\pi$. The positions of the negative
peak and positive peak are determined by the parameters of the
system. In order to illustrate the transport behavior, currents
$j(F_{0})$, $-j(F_{0})$ and $J$ versus $x_{0}$ for $F_{0}=0.5$ are
shown in Fig. 3(b). When $0\leq x_{0}<\frac{\pi}{2}$ and
$\frac{3}{2}\pi<x_{0}\leq 2\pi$, $j(F_{0})$ is more than
$-j(-F_{0})$, so the current $J$ is positive. When
$\frac{\pi}{2}<x_{0}<\frac{3}{2}\pi$, $-j(-F_{0})$ is more than
$j(F_{0})$ and $J$ is negative. $j(F_{0})$ is equal to $-j(-F_{0})$
at $x_{0}=\frac{\pi}{2}$ and $\frac{3}{2}\pi$.

\indent In fact, the sign of the current is determined by the
asymmetry of the tube. The corresponding shape of the tube is given
by Fig. 4. It is obvious that the initial coordinate $x_{0}$
determines the asymmetry of the tube. When the left space of the
tube is bigger than the right ($x_{0}=0$), the particles run to the
right in the presence of the unbiased external force. The particles
move to the left when the right space is bigger ($x_{0}=\pi$). There
is no current for the symmetric tube ($x_{0}=\frac{\pi}{2},
\frac{3}{2}\pi$). Hence, the system can pump the particles from the
left to the right for $0\leq x_{0}<\frac{\pi}{2}$ and
$\frac{3}{2}\pi<x_{0}\leq 2\pi$. We take $x_{0}=0$ in the rest
study.

%\begin{figure}[htbp]
 % \begin{center}\includegraphics[width=10cm,height=8cm]{fig4.eps}
%\caption{The tube shape for different values of $x_{0}$. The tube
%is symmetric at $x_{0}=\pi/2$ and $3\pi/2$. The left space is
%bigger than the right at $x_{0}=0$ and the right space is bigger
%at $x_{0}=\pi$. }\label{1}
%\end{center}
%\end{figure}

\subsection{Ratio $\rho_{1}/\rho_{0}$ at $J=0$}

\indent We now focus on the maximum concentration ratio at both end
of the tube. The analytical results are given by Eq. (11). In Figs.
5-7, we investigate the pumping capacity and see how the
concentration ratio depends on  temperature $T$, the radius $r_{b}$
at the bottleneck and the amplitude $F_{0}$ of the external force.

%\begin{figure}[htbp]
 % \begin{center}\includegraphics[width=10cm,height=8cm]{fig5.eps}
% \caption{\baselineskip 0.4in Ratio of concentrations as a function of $T$ at $a=\frac{1}{2\pi}$, $b=\frac{1.5}{2\pi}$, $\alpha=1/2$, $F_{0}=0.5$, and $x_{0}=0$.}\label{1}
%\end{center}
%\end{figure}
\indent Figure 5 presents the concentration ratio
$\rho_{1}/\rho_{0}$ as a function of temperature $T$ at $x_{0}=0$.
The curve is observed to be bell shaped. When $T\rightarrow 0$, the
particles cannot reach all space of the tube and the effect of the
asymmetric entropic barrier disappears and there is no pumping
capacity ($\rho_{1}/\rho_{0}=1$). When $T\rightarrow \infty$, the
effect of the unbiased external force disappears and the pumping
capacity goes to zero, also. Therefore, there is an optimized value
of $T$ at which the ratio takes its maximum value, which indicates
that the thermal noise can facilitate the particles pumping.

%\begin{figure}[htbp]
 % \begin{center}\includegraphics[width=10cm,height=8cm]{fig6.eps}
 %\caption{\baselineskip 0.4in Ratio of concentrations as a function of $r_{b}$ at $a=\frac{1}{2\pi}$, $\alpha=1/2$, $F_{0}=0.5$, $T=0.5$, and $x_{0}=0$. }\label{1}
%\end{center}
%\end{figure}

\indent The ratio $\rho_{1}/\rho_{0}$ as a function of the radius
$r_{b}$ at the bottleneck is plotted in Fig. 6. When
$r_{b}\rightarrow 0$, no particle can pass through the tube, thus,
$\rho_{1}/\rho_{0}\rightarrow\infty$.  With increasing of the radius
$r_{b}$, the effect of the asymmetric entripic barrier reduces and
the pumping capacity decreases.

%\begin{figure}[htbp]
 % \begin{center}\includegraphics[width=10cm,height=8cm]{fig7.eps}
 %\caption{\baselineskip 0.4in Ratio of concentrations as a function of $F_{0}$ at $a=\frac{1}{2\pi}$, $b=\frac{1.5}{2\pi}$, $\alpha=1/2$, $T=0.5$, and $x_{0}=0$. }\label{1}
%\end{center}
%\end{figure}

\indent Figure 7 shows the ratio $\rho_{1}/\rho_{0}$ vs the
amplitude $F_{0}$ of the unbiased external force. When
$F_{0}\rightarrow 0$, only the effect of the asymmetric entropic
barrier exists, so the system has no pumping capacity. The ratio
$\rho_{1}/\rho_{0}$ saturates to $1$ in the large amplitude $F_{0}$
limit. Therefore, the pumping capacity is maximum when $F_{0}$ takes
an optimized value.

\subsection{Current $J$ at $\rho_{1}>\rho_{0}$}

%\begin{figure}[htbp]
 % \begin{center}\includegraphics[width=10cm,height=8cm]{fig8.eps}
 %\caption{\baselineskip 0.4in Current $J$ vs temperature $T$ for different values of $\rho_{1}$ at $a=\frac{1}{2\pi}$, $b=\frac{1.5}{2\pi}$, $\alpha=1/2$, $F_{0}=0.5$, $x_{0}=0$, and $\rho_{0}=1$.}\label{1}
%\end{center}
%\end{figure}
\indent In Fig. 8, the current $J$ versus temperature $T$ is
presented for different values of $\rho_{1}$ at $\rho_{0}=1$. For
lower values of $\rho_{1}$ the current is larger. When
$\rho_{1}>\rho_{0}$, the current is negative for both too low or too
high temperature. When $T\rightarrow 0$, the effect of the
asymmetric entropic barrier disappears, the concentration difference
dominates the transport, thus, the current is negative. When
$T\rightarrow \infty$, the effect of the unbiased external force
disappears, the transport is dominated by the concentration
difference and the particles move to the left. If the concentration
difference exceed its maximum value, the current will be always
negative.

%\begin{figure}[htbp]
 % \begin{center}\includegraphics[width=10cm,height=8cm]{fig9.eps}
 %\caption{\baselineskip 0.4in Current $J$ as a function of $r_{b}$ at $a=\frac{1}{2\pi}$, $\alpha=1/2$, $F_{0}=0.5$, $T=0.5$, $x_{0}=0$, $\rho_{0}=1$, and $\rho_{1}=4$ }\label{1}
%\end{center}
%\end{figure}

\indent Figure 9 shows the current $J$ as a function of the radius
$r_{b}$ at $\rho_{0}=1$, $\rho_{1}=4$. The curve is observed to be
bell shaped. When $r_{b}\rightarrow 0$, the tube is jammed and the
particles cannot pass through the tube. When
$r_{b}\rightarrow\infty$, the effect of the asymmetric entropic
barrier disappears, and the current is dominated by the
concentration difference, thus the current is negative. So there
exists an optimized value of $r_{b}$ at which the current takes its
maximum value.

\section{Concluding Remarks}
\indent In this study, we study the transport of Brownian
particles moving in an asymmetric finite tube in the presence of
an unbiased external force. The pumping device is embedded in a
finite region and bounded by two particle reservoirs. It is found
that the particles can be pumped from low concentration reservoir
to higher concentration reservoir. When $\rho_{1}=\rho_{0}$, the
current is positive for $0<x_{0}<\pi/2$ and $3\pi/2<x_{0}<2\pi$,
zero at  $x_{0}=\pi/2$ and $3\pi/2$, negative for
$2\pi<x_{0}<3\pi/2$. The sign of the current is determined by the
asymmetry of the tube. The maximum ratio of the concentrations
($J=0$) at both ends of the tube is also studied. We can find that
there is an optimized value of temperature $T$ (or the amplitude
$F_{0}$ of the external force) which gives the maximum ratio of
$\rho_{1}/\rho_{0}$. The ratio $\rho_{1}/\rho_{0}$ decreases with
increasing the radius $r_{b}$ at the bottleneck of the tube.  We
also study the current $J$ as a function of $J$ and $r_{b}$. We
observe that the current can tend to its maximum value when $T$ or
$r_{b}$ take the optimized value.

\indent Though the model presented does not pretend to be a
realistic model for a biological pumps, the results we have
presented have a wide application in many systems, such as Na,
K-ATPase pumps \cite{17}, membrane proteins \cite{18}, and electron
pumps \cite{19}.
\begin{center}
    \textbf{{ACKNOWLEDGMENTS}}
\end{center}
 \indent The work was supported by the National
Natural Science Foundation of China under Grant No. 30600122 and
GuangDong Provincial Natural Science Foundation under Grant No.
06025073.

\newpage
\section{Caption List}
\baselineskip 0.4in
 Fig. 1. Scheme of the pumping device: a
spatially asymmetric tube is embedded in a finite region of length
$L$ and bounded by two particle reservoirs of concentrations
$\rho_{0}$ and $\rho_{1}$. The shape of the tube is determined by
its radius $\omega(x)$ and the left end coordinate $x_{0}$ of the
tube. The particles in the
tube are powered by an unbiased external force $F(t)$.\\

Fig.2.  Current $J$ as a function of period numbers $N$ at
$x_{0}=0$, $a=\frac{1}{2\pi}$, $b=\frac{1.5}{2\pi}$, $\alpha=1/2$,
$F_{0}=0.5$, $T=0.5$, $\rho_{0}=1$, and $\rho_{1}=1$. \\

Fig. 3. (a) Current $J$ as a function of $x_{0}$ for different
values of $F_{0}$ at $a=\frac{1}{2\pi}$, $b=\frac{1.5}{2\pi}$,
$\alpha=1/2$, $T=0.5$, $\rho_{0}=1$, and $\rho_{1}=1$. (b)Currents
$j(F_{0})$, $-j(-F_{0})$ and $J$ versus $x_{0}$ for $F_{0}=0.5$.
The other parameters are the same as that in (a).\\

Fig. 4. The tube shape for different values of $x_{0}$. The tube
is symmetric at $x_{0}=\pi/2$ and $3\pi/2$. The left space is
bigger than the right at $x_{0}=0$ and the right space is bigger
at $x_{0}=\pi$. \\

Fig. 5. Ratio of concentrations as a function of $T$ at
$a=\frac{1}{2\pi}$, $b=\frac{1.5}{2\pi}$, $\alpha=1/2$,
$F_{0}=0.5$, and $x_{0}=0$.\\

Fig. 6. Ratio of concentrations as a function of $r_{b}$ at
$a=\frac{1}{2\pi}$, $\alpha=1/2$, $F_{0}=0.5$, $T=0.5$, and
$x_{0}=0$.\\

Fig. 7. Ratio of concentrations as a function of $F_{0}$ at
$a=\frac{1}{2\pi}$, $b=\frac{1.5}{2\pi}$, $\alpha=1/2$, $T=0.5$,
and $x_{0}=0$.\\

Fig. 8. Current $J$ vs temperature $T$ for different values of
$\rho_{1}$ at $a=\frac{1}{2\pi}$, $b=\frac{1.5}{2\pi}$,
$\alpha=1/2$, $F_{0}=0.5$, $x_{0}=0$, and $\rho_{0}=1$.\\

Fig. 9. Current $J$ as a function of $r_{b}$ at
$a=\frac{1}{2\pi}$, $\alpha=1/2$, $F_{0}=0.5$, $T=0.5$, $x_{0}=0$,
$\rho_{0}=1$, and $\rho_{1}=4$.\\

\newpage
\begin{figure}[htbp]
\begin{center}\includegraphics[width=10cm,height=8cm]{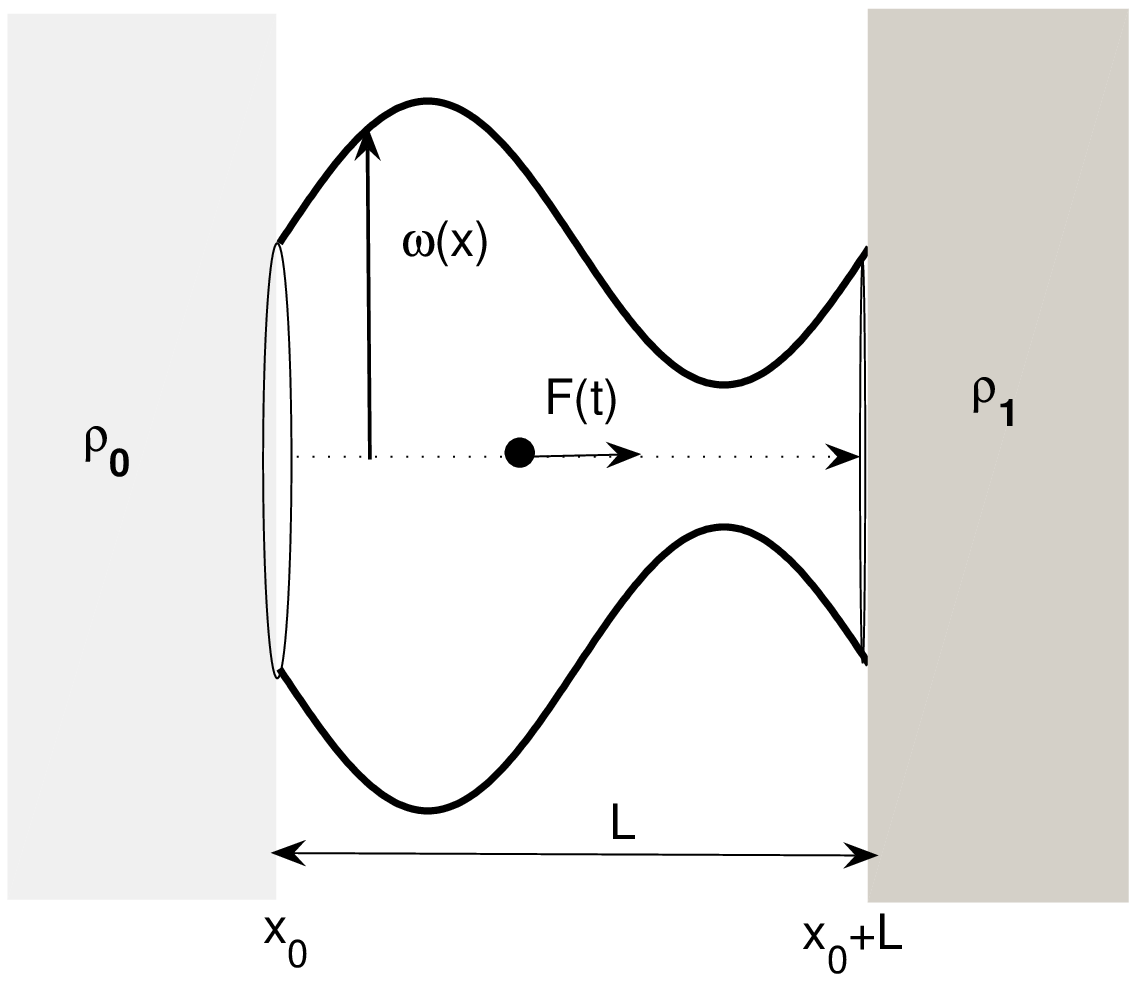}
 \caption{}\label{1}
\end{center}
\end{figure}

\newpage
\begin{figure}[htbp]
\begin{center}\includegraphics[width=10cm,height=8cm]{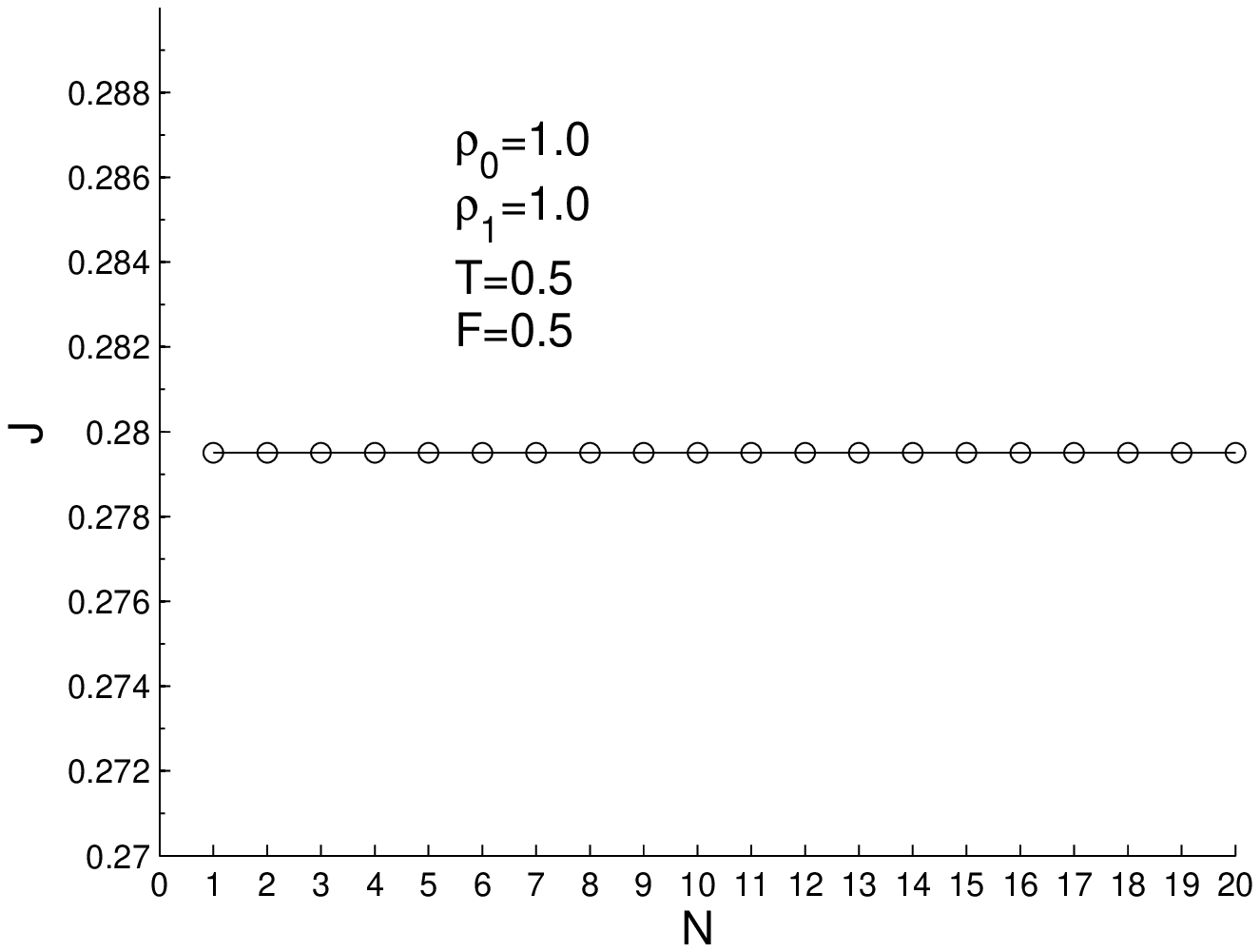}
 \caption{}\label{1}
\end{center}
\end{figure}

\newpage
\begin{figure}[htbp]
\begin{center}\includegraphics[width=10cm,height=8cm]{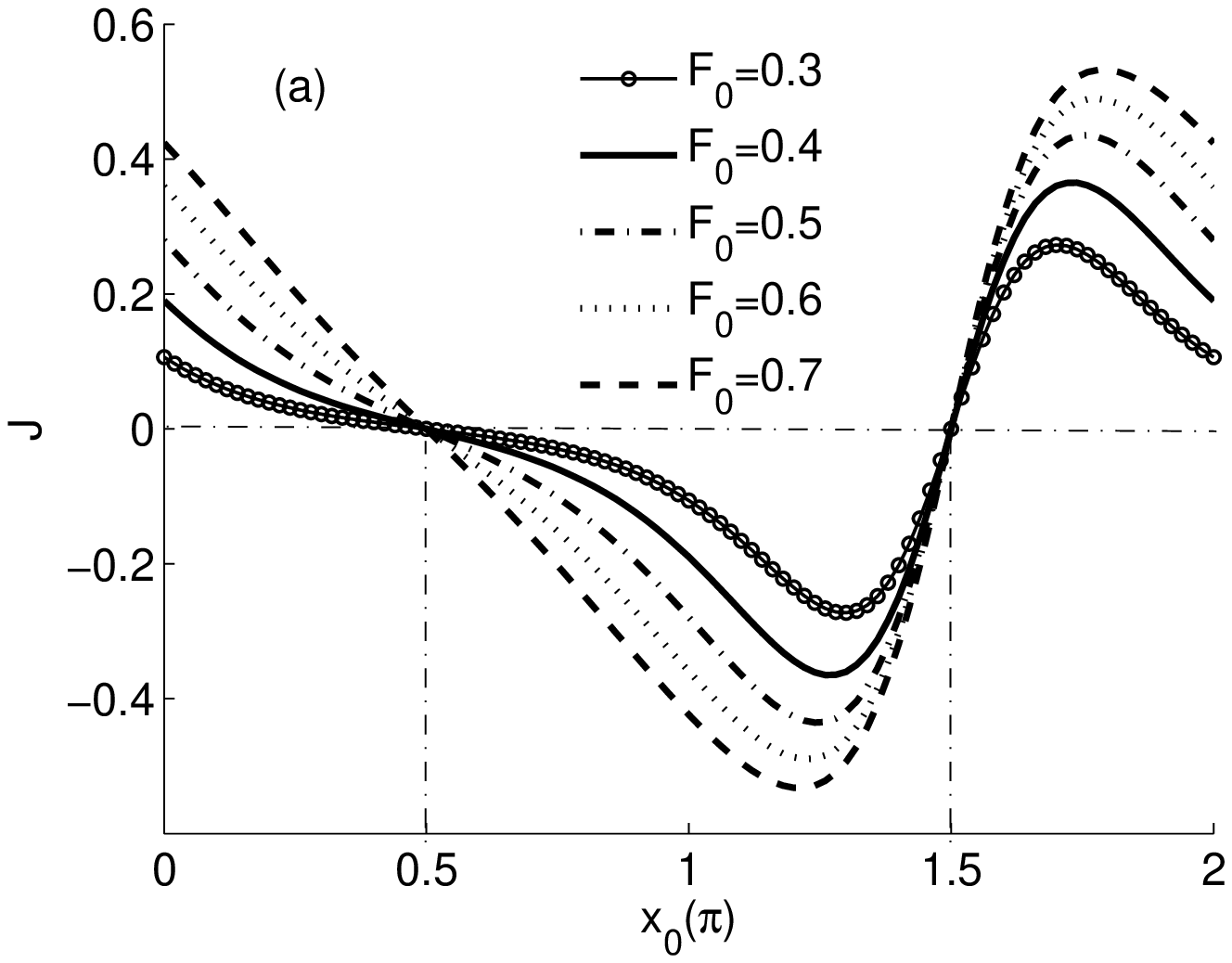}
\includegraphics[width=10cm,height=8cm]{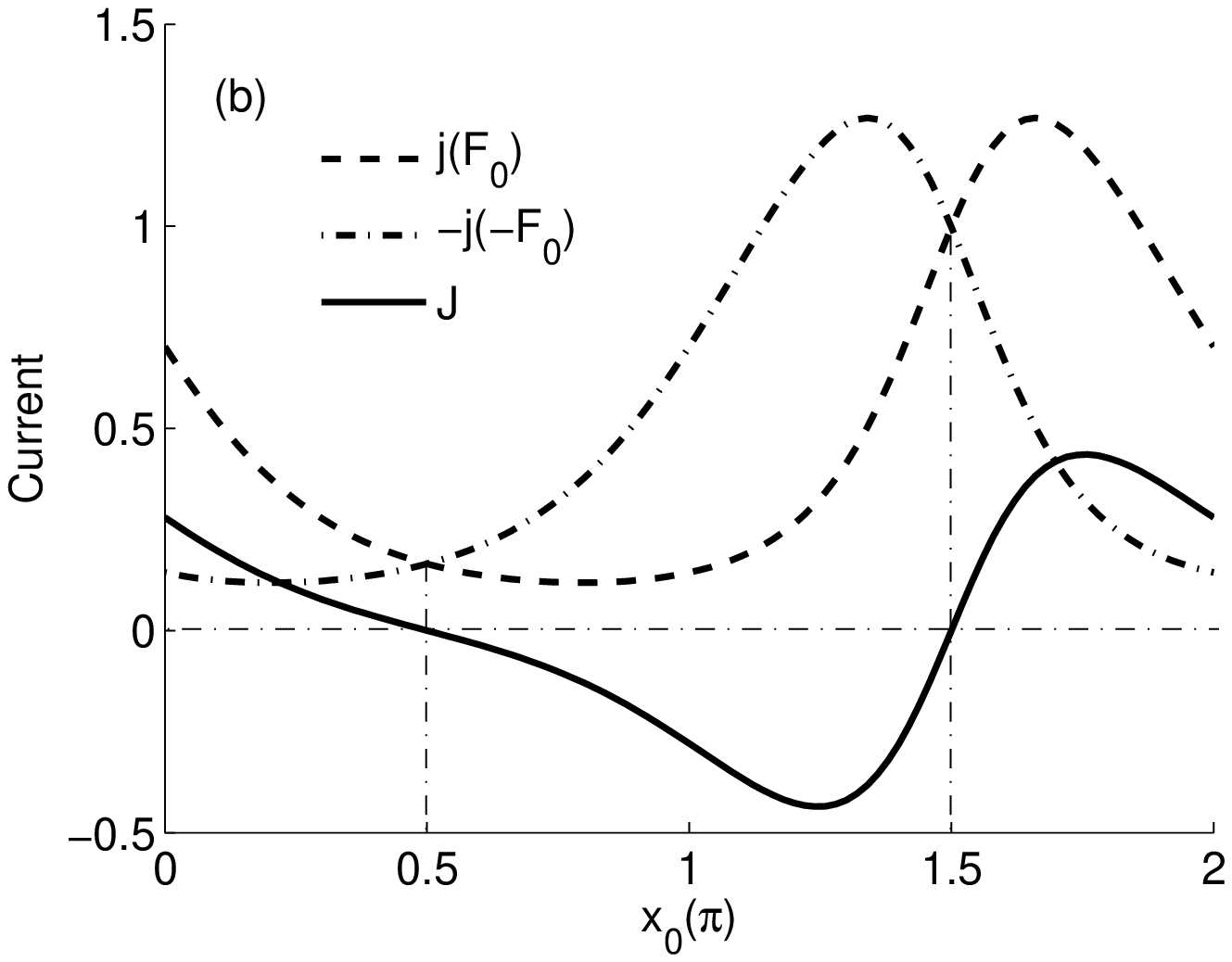}
 \caption{}\label{1}
\end{center}
\end{figure}

\newpage
\begin{figure}[htbp]
\begin{center}\includegraphics[width=10cm,height=8cm]{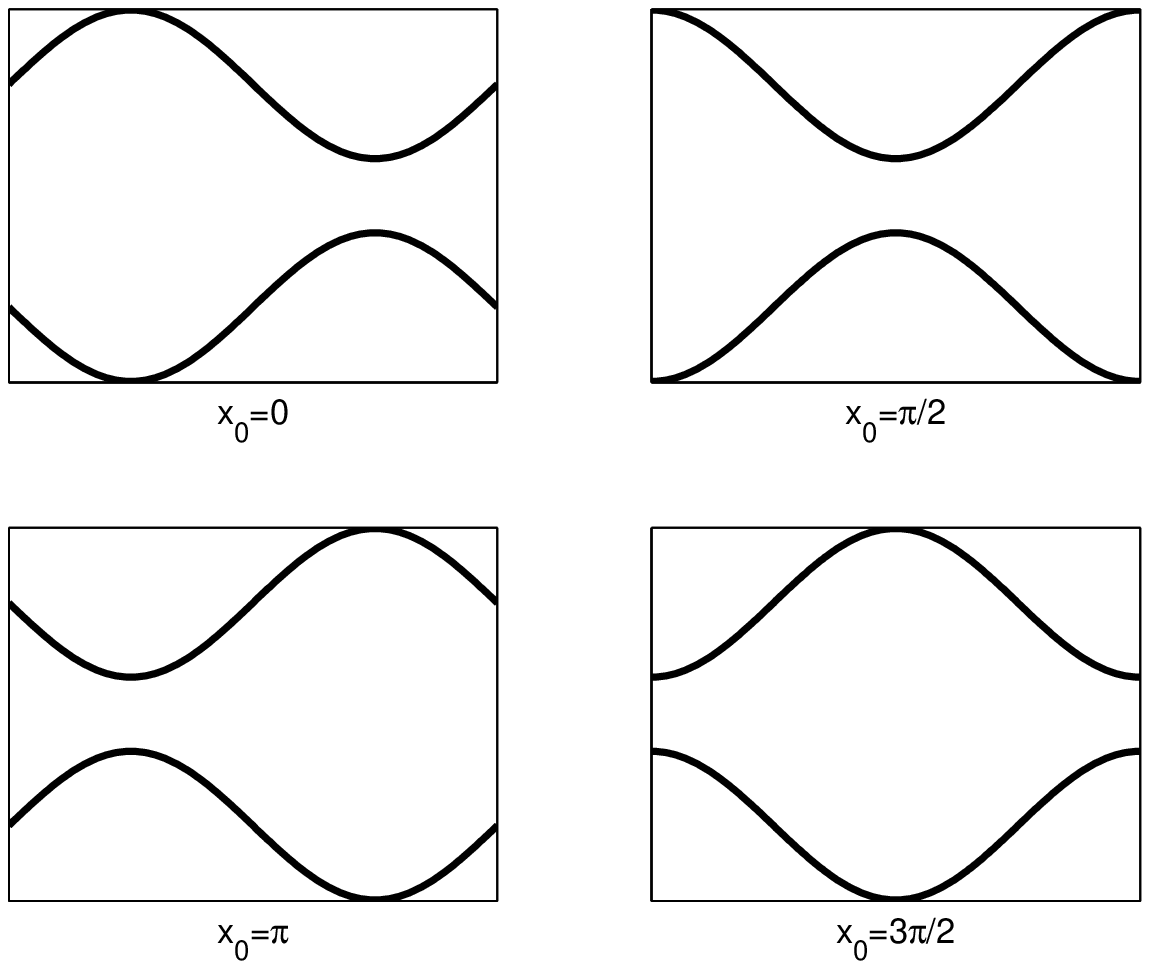}
 \caption{}\label{1}
\end{center}
\end{figure}

\newpage
\begin{figure}[htbp]
\begin{center}\includegraphics[width=10cm,height=8cm]{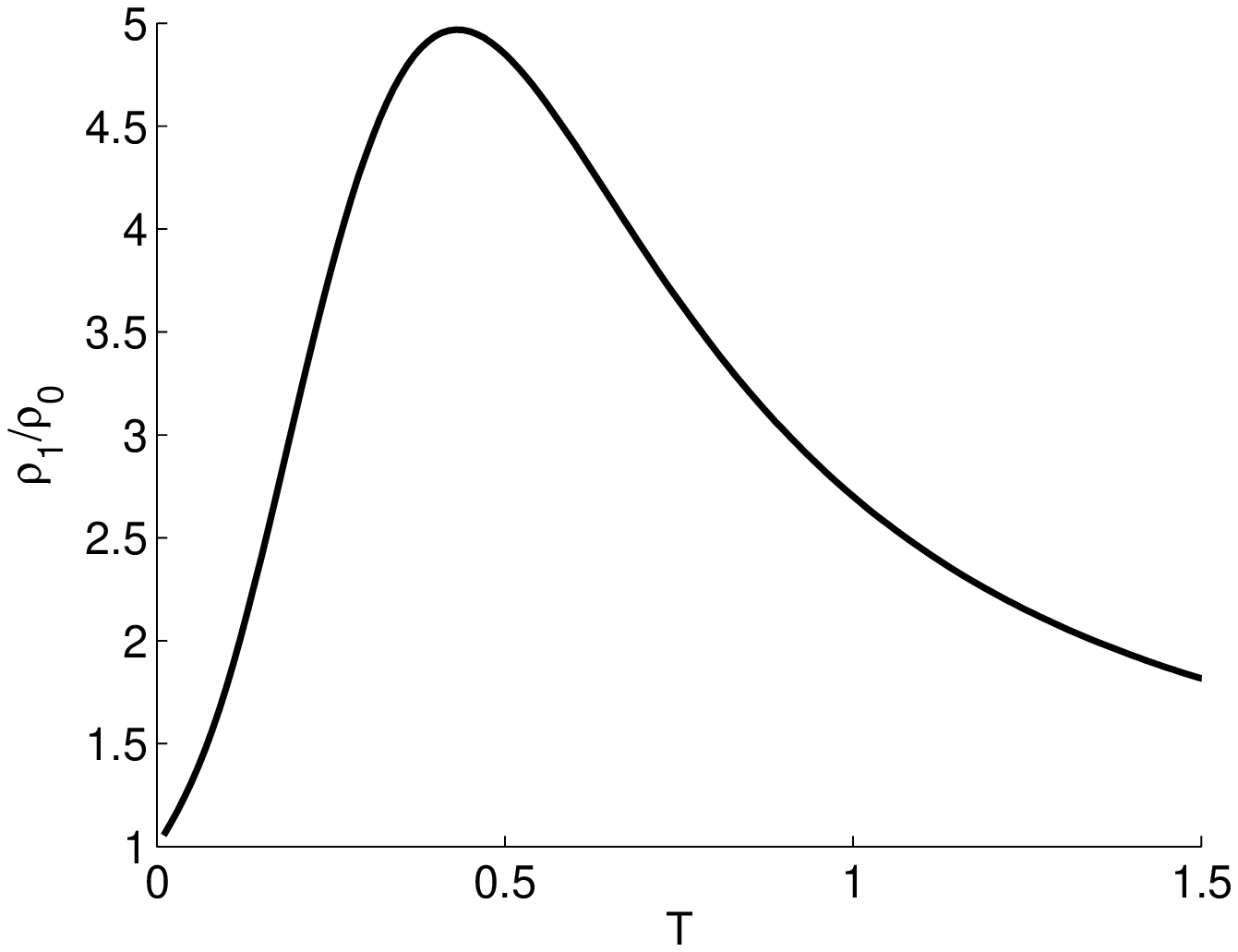}
 \caption{}\label{1}
\end{center}
\end{figure}

\newpage
\begin{figure}[htbp]
\begin{center}\includegraphics[width=10cm,height=8cm]{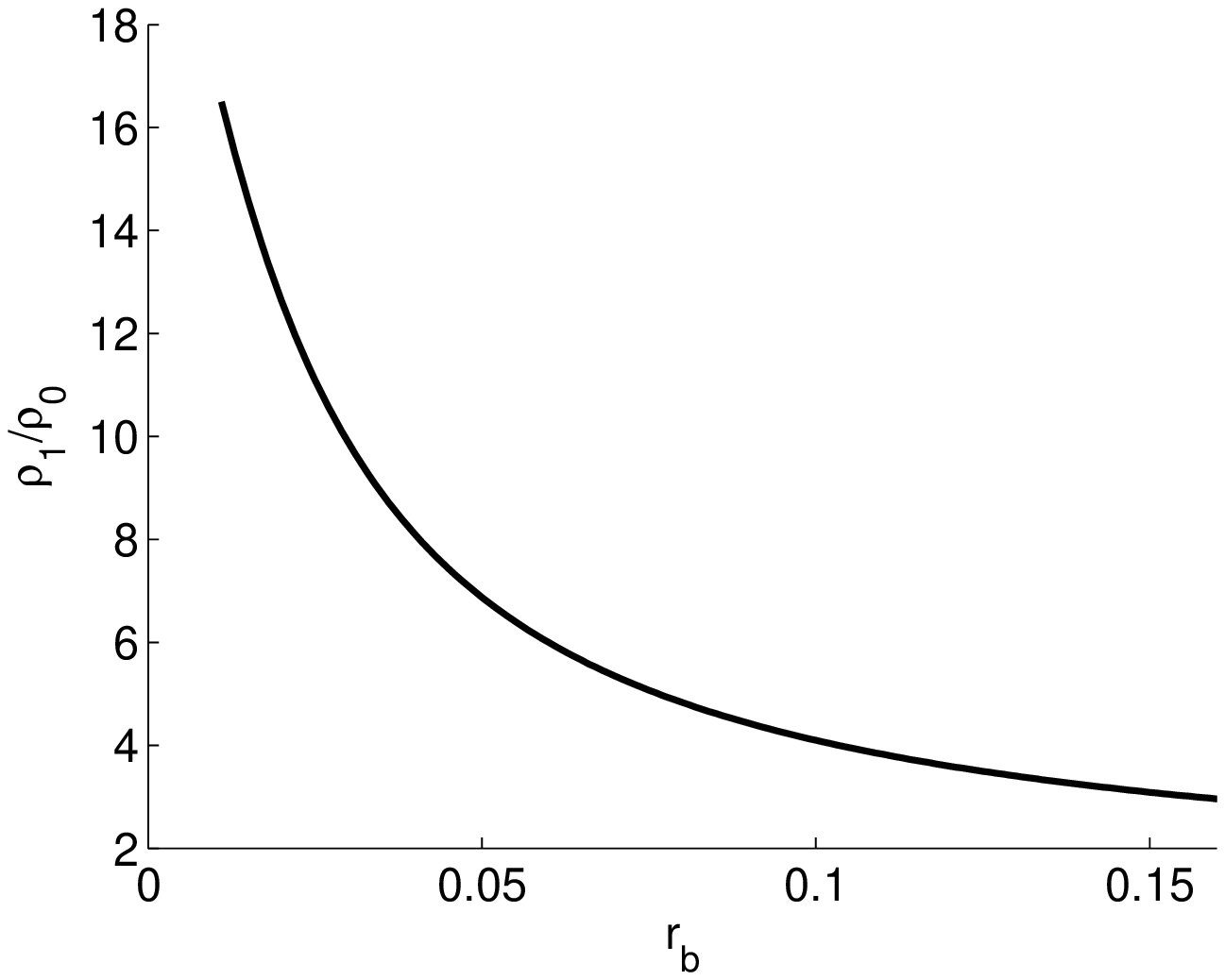}
 \caption{}\label{1}
\end{center}
\end{figure}

\newpage
\begin{figure}[htbp]
\begin{center}\includegraphics[width=10cm,height=8cm]{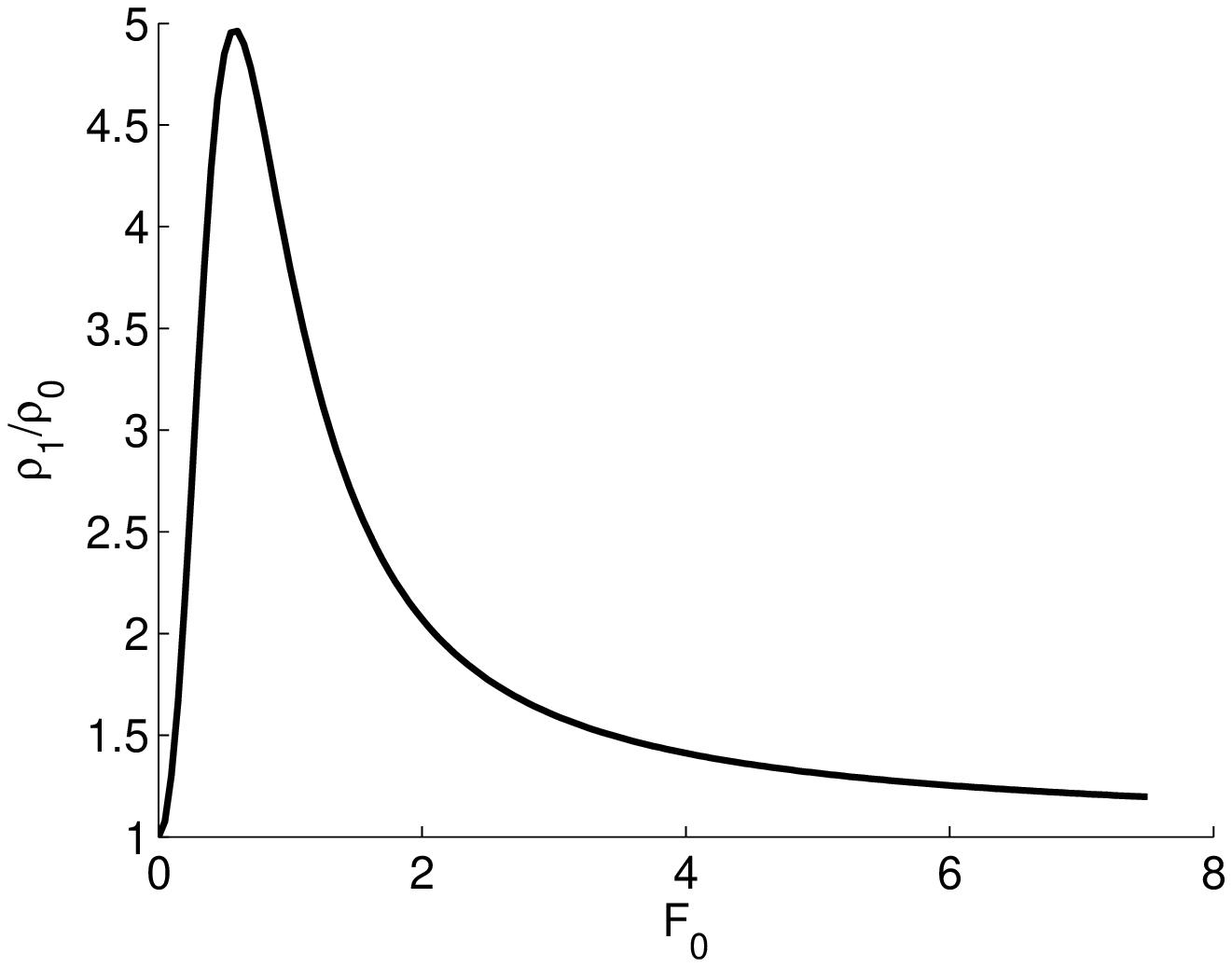}
 \caption{}\label{1}
\end{center}
\end{figure}

\newpage
\begin{figure}[htbp]
\begin{center}\includegraphics[width=10cm,height=8cm]{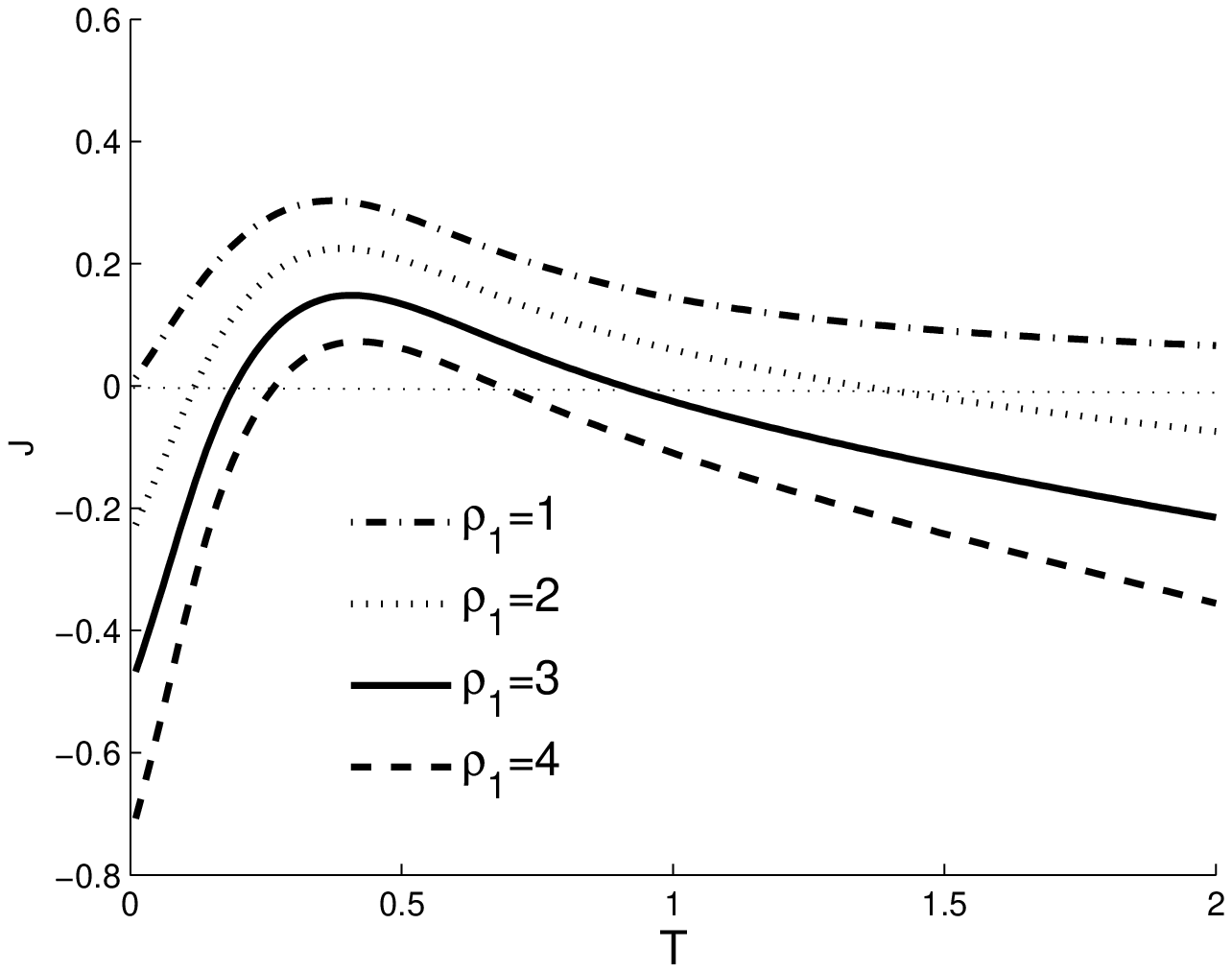}
 \caption{}\label{1}
\end{center}
\end{figure}

\newpage
\begin{figure}[htbp]
\begin{center}\includegraphics[width=10cm,height=8cm]{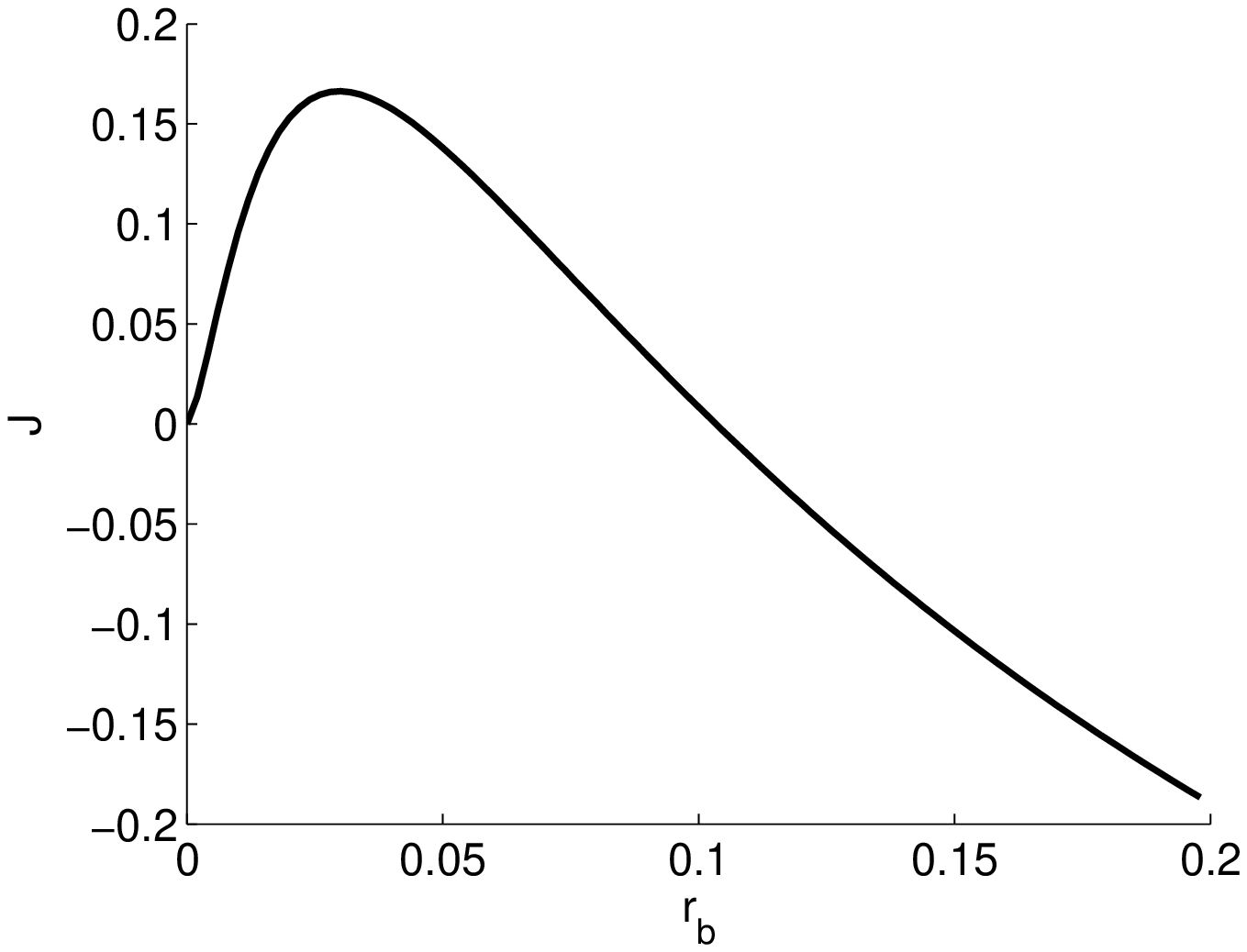}
 \caption{}\label{1}
\end{center}
\end{figure}


\begin{thebibliography}{}
\baselineskip 0.4in
\bibitem{1}P. Reimann, Phys. Rep. 361, 57 (2002).
\bibitem{2}R. D. Astumian and P. Hanggi, Physics Today 55, 33(2002).
\bibitem{3}P. Hanggi, F. Marchesoni, and F. Nori, Ann. Phys. (Leipzip)14, 51 (2005).
\bibitem{4}R. P. Feynman, R. B. Leighton and M. Sands, The Feynman
lectures on physics (Addison Wesley, Reading, MA, 1963), Vol. 1,
pp.46.1-46.9.
\bibitem{5}J. Prost, J. F. Chauwin, L. Peliti, and A. Ajdari, Phys.
Rev. Lett. 72, 2652 (1994).
\bibitem{6}R. D. Astumian and I. Derenyi, Phys. Rev. Lett. 86,
3859(2001); R. D. Astumian, Phys. Rev. Lett. 91, 118102 (2003).
\bibitem{7}I. Kosztin and K. Schulten, Phys. Rev. Lett. 93, 238102
(2004).
\bibitem{8}M. Rey, M. Strass, S. Kohler, P. Hanggi, and F. Sols,
Phys. Rev. B 76, 085337 (2007).

\bibitem{9}J. F. Wambaugh, C. Reichhardt, C. J. Olson, F. Marchesoni, and F. Nori, Phys. Rev. Lett. 83, 5106 (1999).
\bibitem{10}S. Savelev, V. Misko, F. Marchesoni, and F. Nori, Phys. Rev. B 71, 214303 (2005).
\bibitem{11}J. M. Sancho and A. Gomez-Marin, Europhysics Letters 86, 40002 (2009); J. M. Sancho and A. Gomez-Marin, Phys. Rev. E
77, 031108(2008).
\bibitem{12}D. Reguera and J. M. Rubi, Phys. Rev. E 64, 061106
(2001); D. Reguera, G. Schmid, P. S. Burada, J. M. Rubi, P. Reimann,
and P. Hanggi, Phys. Rev. Lett. 96, 130603 (2006).
\bibitem{13}R. Zwanzig, J. Phys. Chem. 96, 3926 (1992).
\bibitem{14}P. Kalinay and J. K. Percus, Phys. Rev. E 74, 041203 (2006).
\bibitem{15}B. Q. Ai and L. G. Liu, Phys. Rev. E 74, 051114 (2006);
B. Q. Ai, H. Z. Xie, and L. G. Liu, Phys. Rev. E 75, 061126 (2007);
B. Q. Ai, L. G. Liu, J. Chem. Phys. 126, 204706 (2007).
\bibitem{16}A. M. Berezhkovskii, M. A. Pustovoit, and S. M.
Bezrukov, J. Chem. Phys. 126, 134706 (2007).
\bibitem{17}D. S. Liu, R. D. Astumian, and T.Y. Tsong, J. Biol. Chem.
265, 7260 (1990); T. D. Xie, P. Marszalek, Y. D. Chen, and T.Y.
Tsong, Biophys. J. 67, 1247 (1994); T. D. Xie, Y. D. Chen, P.
Marszalek, and T.Y. Tsong, Biophys. J. 72, 2496 (1997).

\bibitem{18}S. Ramaswamy, J. Toner, and J. Prost, Phys. Rev. Lett. 84,
3494 (2000); A. E. Pelling et al., Science 305, 1147 (2004);M.
Borgnia, S. Nielsen, A. Engel, and P. Agre, Annu. Rev. Biochem.
68, 425 (1999).
\bibitem{19}M. Switkes, C. M. Marcus, K. Campman, and A. C.
Gossard, Science 283, 1905 (1999).
\end{thebibliography}
\end{document}